\documentstyle[preprint,aps]{revtex}
\begin{document}
\title{Tight-binding parametrization of first-principles
electronic dispersion in orientationally disordered $A_3$C$_{60}$}
\author{S.C. Erwin and E.J. Mele}
\address{Department of Physics and Laboratory for Research on the Structure
of Matter\\University of Pennsylvania, Philadelphia, Pennsylvania 19104}
\date{January 24, 1994}
\maketitle
\begin{abstract}
We derive numerical tight-binding hopping parameters to describe
conduction-band dispersion in arbitrary orientational phases of
$A_3$C$_{60}$. The parameters are obtained by direct Fourier inversion
of the spectra from self-consistent electronic-structure
calculations for K$_3$C$_{60}$ using the local-density approximation,
including the effects of orientational dependence. Using the new
parameters, we revisit several earlier investigations of the
orientational ordering in $A_3$C$_{60}$; some of the earlier results
are substantiated, while others are slightly modified.

\end{abstract}


The electronic structure of the fullerene and fulleride solids has
been treated at various levels of theoretical sophistication,
including simple H\"uckel calculations for the isolated molecule,
fully self-consistent density-functional methods using the
local-density approximation (LDA) for the intercalated solids, and
most recently, quasiparticle-corrected $GW$ methods for solid C$_{60}$
\cite{haddon,pickett,shirley}.  For the $A_x$C$_{60}$ family
($x$=0,3,4,6, $A$=heavy alkali), a single common picture emerges:
C$_{60}$ molecular states are broadened into band states in the solid,
with bandwidths on the order of $\sim$1 eV or less, and the alkali
$s$-electrons populate the C$_{60}$-derived band states.  The small
dispersion, together with the one-to-one correspondence between
molecular levels and bands, implies a nearly negligible redistribution
of electronic charge density on forming the condensed phase (the
alkali ionization aside). This suggests that a tight-binding (TB)
Hamiltonian might provide an accurate description of dispersion in
this family, that TB parameters should be transferable among similar
structures, and that interactions may be limited to nearest neighbors.

We focus on the superconducting $A_3$C$_{60}$ phase because the nature
of fullerene molecular orientational order in this phase is not well
understood, having been variously described as merohedrally disordered
with $Fm\overline{3}m$ symmetry \cite{stephens}; orientationally
disordered with lower symmetry \cite{barrett}; and partially ordered on
short length scales \cite{egami}.  Important theoretical work by
Gelfand and Lu \cite{gelfand1} and by Satpathy {\it et al.}
\cite{satpathy} demonstrated that complete merohedral disorder has a
profound effect on the conduction-band spectral density and on ac
conductivity, effectively washing out the sharp structure that
normally results from a well-defined band structure.  Deshpande {\it
et al.} \cite{deshpande} subsequently developed an effective-medium
theory to describe the effects of disorder in a Green's function
formalism, essentially confirming the results obtained numerically by
Gelfand and Lu. Recently, Mele and Erwin \cite{mele1} used both formal
and numerical methods to show that although disorder strongly scatters
electron states and substantially redistributes spectral density,
states near the Fermi energy are characterized by an effective Bloch
wave vector and a mean free path of order 2--3 near-neighbor
spacings. Consequently, a relatively well-defined ``Fermi surface''
may be considered for these states even in the presence of random
(merohedral) disorder. In all of these studies, the Hamiltonian of
Ref.~\cite{gelfand1} was used, for which hopping parameters were
calculated from C$_{60}$ interactions at the level of H\"uckel
theory. In this Report, we provide an explicit parametrized
Hamiltonian for $A_3$C$_{60}$, with TB parameters derived directly
from first-principles LDA calculations of dispersion in $A_3$C$_{60}$,
with the effects of orientational disorder fully accounted for.

The structure and symmetry of the
nearest-neighbor hopping
Hamiltonian have been discussed in detail by Yildirim {\it et al.}
\cite{yildirim1}, and we will adopt their notation for the discussion
here. The Hamiltonian describing conduction-band dispersion is
\begin{equation}
{\cal H} = \sum_{{\bf R},{\bf \tau}} \sum_{\alpha\beta}
t_{\alpha\beta}({\bf \tau};\sigma_{\bf R},\sigma_{{\bf R}+{\bf \tau}})
c^\dagger_{{\bf R},\alpha} c_{{\bf R}+{\bf \tau},\beta},
\label{hopham}
\end{equation}
where $\alpha,\beta=x,y,z$ is the orbital polarization,
$\sigma_{\bf R}=\pm 1$ is an Ising
variable giving the orientation of the molecule at ${\bf R}$, and
${\bf \tau}=(a/2)(110)$, etc., gives the direction of the hop.
The hopping matrix has four possible values, and can be conveniently
decomposed as:
\begin{eqnarray}
\lefteqn{t_{\alpha\beta}({\bf \tau};
\sigma_{\bf R},\sigma_{{\bf R}+{\bf \tau}})=}
\nonumber\\
&&t^{(0)}_{\alpha\beta}({\bf \tau})+
t^{(1)}_{\alpha\beta}({\bf \tau})\sigma_{\bf R}+
t^{(2)}_{\alpha\beta}({\bf \tau})\sigma_{{\bf R}+{\bf \tau}}+
t^{(3)}_{\alpha\beta}({\bf \tau})\sigma_{\bf R}\sigma_{{\bf R}+{\bf \tau}}.
\label{fourterms}
\end{eqnarray}
For the hopping direction ${\bf \tau}_0\equiv (a/2)(110)$, Yildirim
{\it et al.} showed from symmetry considerations that the $t^{(i)}$
take the form
\[
t^{(0)}({\bf \tau}_0) =
t \left[ \begin{array}{rrr}A&B&0\\B&A&0\\0&0&C\end{array} \right],
t^{(1)}({\bf \tau}_0) =
t \left[ \begin{array}{rrr}X&Y&0\\-Y&-X&0\\0&0&0\end{array} \right],
\]
\begin{equation}
t^{(2)}({\bf \tau}_0) =
t \left[ \begin{array}{rrr}X&-Y&0\\Y&-X&0\\0&0&0\end{array} \right],
t^{(3)}({\bf \tau}_0) =
t \left[ \begin{array}{rrr}D&E&0\\E&D&0\\0&0&F\end{array} \right].
\label{tmats-yil}
\end{equation}
A different, equivalent linear combination of these eight
constants was used by Gelfand and Lu, who defined hopping matrices
between ``parallel'' (like orientation) and ``perpendicular'' (unlike
orientation) molecules, within a basis that rotates with the
molecules.  These two matrices have the form
\begin{equation}
t({\bf \tau}_0)_\| =
t \left[ \begin{array}{rrr}a&b&0\\b&c&0\\0&0&d\end{array} \right],
t({\bf \tau}_0)_\bot =
t \left[\begin{array}{rrr}e&f&0\\-f&g&0\\0&0&h\end{array} \right].
\label{tmats-gel}
\end{equation}

To fit the constants in Eqs.~(\ref{tmats-yil}) and (\ref{tmats-gel})
to self-consistent LDA spectra, we first
consider a supercell of the $A_3$C$_{60}$ fcc Bravais lattice, with a
basis of C$_{60}$ molecules with orientations $\sigma_{{\bf R}_i}$
$(i=1,N)$.  If the orientations are not all identical, then all four
terms from Eq.~(\ref{fourterms}) will contribute to $\cal H$. By
calculating the LDA eigenvalues, $E^{LDA}_{n{\bf k}}$, throughout the
zone, optimal values (in the least-squares sense) for all of the
constants in Tables I and II can be determined.  We start by assuming we
know the eigenvalues, $E^{TB}_{n{\bf k}}$, and eigenvectors, $|n{\bf
k}\rangle$, of $\cal H$ in the TB basis for a particular set of hopping
constants. Then ${\cal H}({\bf k})$ can be written as
\begin{equation}
{\cal H}({\bf k}) = \sum_n  E^{TB}_{n{\bf k}} |n{\bf k}\rangle
\langle n{\bf k}|.
\label{hk}
\end{equation}
{}From Eq.~(\ref{hopham}), the matrix elements of ${\cal H}({\bf k})$ in the TB
basis are also given explicitly in terms of the hopping parameters:
\begin{equation}
\langle {\bf R},\alpha | {\cal H}({\bf k}) | {\bf R}+{\bf
\tau},\beta\rangle = \sum_{\bf \tau}
t_{\alpha\beta}({\bf \tau};\sigma_{\bf R},\sigma_{{\bf R}+{\bf \tau}})
\exp(i {\bf k}\cdot{\bf \tau}).
\label{htb}
\end{equation}
Eqs.~(\ref{hk}) and (\ref{htb}) are equivalent formulations. We
exploit this by first constructing ${\cal H}({\bf k})$ according to
Eq.~(\ref{hk}) but with the substitution $E^{TB}_{n{\bf k}}
\rightarrow E^{LDA}_{n{\bf k}}$. The resulting Hamiltonian matrix
is Fourier inverted according to Eq.~(\ref{htb}) to yield optimal
hopping parameters in the least-squares sense. In practice, the
solution is found iteratively, with the hopping constants from
Ref.~\cite{gelfand1} providing the seed eigenvectors.

The supercell we choose is the smallest possible, containing two
C$_{60}$ molecules (with different orientations) and six K ions. The
resulting crystal structure has $D_{4h}$ symmetry, and can be
described as sheets of C$_{60}$ molecules alternating along the (001)
direction between the ``$A$'' and ``$B$'' orientations; this
corresponds to the $A_2B_2$ structure of Yildirim {\it et al.}
\cite{yildirim1}. The computational methods for solving the Kohn-Sham
equations are identical to those used in previous studies of
K$_6$C$_{60}$, K$_3$C$_{60}$, and K$_4$C$_{60}$ \cite{erwin}.  The
charge density and potential are completely general and without shape
approximation, and were iterated to self-consistency using the
$\Gamma$ point.  Bloch basis functions are linear combinations of
occupied and unoccupied atomic orbitals for potassium and carbon;
these are in turn expanded on a set of gaussian functions. This
provides a compact basis, and allows us to perform accurate
all-electron calculations---without the need for pseudopotentials---in
which core, valence, and conduction states are treated on equal
footing. The Ceperly-Alder exchange-correlation functional was
used. For the inversion of Eq.~(\ref{htb}), we used a mesh of 40
equally spaced ${\bf k}$-points in the irreducible wedge of the folded
zone.

The hopping constants obtained by fitting to the LDA spectrum (denoted
TB/LDA) are listed in the first row of Tables I and II; the
constants obtained by Gelfand and Lu \cite{gelfand1} (denoted TB/GL) are
listed in the second row of Tables I and II.  The TB/LDA values follow
the qualitative trends in the TB/GL values, but substantial
differences are apparent. To clarify the significance of these
differences, we have used the Hamiltonian of Eq.~(\ref{hopham}) to
compute spectra for three different structures: the orientationally
ordered crystal (Fig.~1); the $A_2B_2$ supercell used to extract the
hopping parameters (Fig.~2); and an ensemble of 27-molecule supercells
with quenched disorder, which simulates the merohedrally disordered
structure originally proposed by Stephens \cite{stephens}
(Fig.~3). For the first two structures, we also compare the TB spectra
to the full self-consistent LDA spectra. We only show
densities-of-states (DOS), since differences are still evident in
these ${\bf k}$-integrated spectra.

Not surprisingly, for the orientationally ordered crystal, the TB/LDA
parameter set reproduces the full LDA calculation more faithfully than
the TB/GL set.  More interesting is the DOS (Fig.~1, middle panel,
light curve) using the hopping constants fit to the LDA
spectrum from the single-orientation fcc crystal.  The close
resemblance to the spectrum obtained by fitting to the LDA spectrum of
the $A_2B_2$ structure illustrates the transferability of a
particular hopping Hamiltonian from one crystal structure to
another. This is the numerical justification for using hopping
constants obtained from the $A_2B_2$ structure to compute spectra
for supercells simulating merohedral disorder.  Interestingly, the
ensemble average of these supercell spectra (Fig.~3) shows only small
differences arising from the two parameter sets.  The small dip at
0.22 eV in the TB/GL data is absent in the TB/LDA
spectrum, and the band edges of the latter show more gradual tailing
than the former; otherwise the spectra are very similar.

Deshpande {\it et al.} \cite{deshpande} have discussed the fact that
the magnitudes and signs of the constants in Tables I and II depend on
the choice of local basis sets for two neighboring molecules. They
defined a simple gauge-invariant quantity by considering the spectrum
of the Hamiltonian for hopping around the smallest possible
closed path, i.e. forward and backward hopping across a single
``bond''. There are two possibilities---hopping between like ($AA$)
and unlike ($AB$) orientations---which the authors showed lead to
numerically similar spectra, as shown in the second row of Table III.
If it were true that hopping proceeded via nearest-neighbor {\it
atomic} sites only, then the structure of the C$_{60}$ molecule and the
crystallographic alignment of the two C$_{60}$ orientations would lead
to exactly degenerate single-bond spectra. The deviation of the values
in Table III from this equality is a direct measure of the importance
of hopping between second-neighbor (and higher) atomic sites. In the
first row of Table III, we have recomputed the single-bond spectra
based on the new hopping constants. Again, changes are evident
(especially for the smallest eigenvalue), but the approximate
degeneracy is still observed. In particular, the largest $AA$ and $AB$
eigenvalues---which dominate the spectral features for all such
retraceable paths---are nearly equal for both sets of constants. This
justifies numerically the treatment of Ref.~\cite{deshpande}, in which
all sites belonging to retraceable paths in the effective medium were
assigned the same orientation.

Finally, we reconsider the results of Yildirim {\it et al.}, who have
made a systematic analysis of the ground-state energetics of various
orientational phases of $A_3$C$_{60}$ \cite{yildirim1}. The authors
considered the Ising-like nature of the orientational ordering
problem, and showed that while the direct Coulomb interaction between
molecules favors a ``ferromagnetic'' ordering, the indirect kinetic
energy contribution from the conduction electrons favors an
antiferromagnetic alignment of nearest neighbors. The orientational
dependence of the Coulomb interaction was analyzed in detail in
Ref.~\cite{yildirim2}; this analysis was based directly on a multipole
expansion of the electronic charge density calculated within LDA, and
so remains unaffected by the present work. The kinetic energy
contribution was analyzed in Ref.~\cite{yildirim1} by calculating the
energy of the filled Fermi sea, $E_{KE}=2\sum_{n{\bf k}} E^{TB}_{n{\bf
k}} f_{n{\bf k}}$, for three different orientational phases: the
orientationally ordered fcc crystal, denoted by $A_4$; the $A_2B_2$
structure discussed above; and a cubic ($Pm\overline{3}$) structure
denoted $A_3B$. In Table IV, we have recomputed $E_{KE}$ for these
three structures, using our TB/LDA hopping constants, as well the
TB/GL constants originally used in Ref.~\cite{yildirim1}.

Several aspects of the results in Table IV deserve comment: (1) Both
sets of hopping parameters favor the non-ferromagnetic $A_3B$ and
$A_2B_2$ configurations; (2) the TB/LDA set predicts the
non-ferromagnetic configurations to be considerably {\it more} favored
than does the TB/GL set; (3) the TB/LDA set predicts the energy
ordering of the $A_3B$ and $A_2B_2$ structures to be opposite to the
prediction of the TB/GL set; (4) the ratio of relative energies,
$\Delta(A_2B_2)/\Delta(A_3B)$, has the value 1.06 for the TB/LDA set,
considerably closer to the ideal nearest-neighbor Ising value of 4/3
than is the TB/GL value of 0.81. This last point suggests that the
TB/LDA kinetic energy contribution to the orientational potential is
{\it also} well modeled by a nearest-neighbor Ising model. When both
Coulomb and kinetic energies are considered, the new parameter set
implies that: (5) Non-ferromagnetic configurations are still favored
so long as the kinetic energy scale, $t$, and the Ising interaction
strength, $J_D$, satisfy the inequality $t>-1.86J_D$.  This condition
is easily satisfied for all values of $J_D$ considered in
Ref.~\cite{yildirim1} (note that the ferromagnetic Coulomb interaction
implies $J_D<0$). (6) Finally, the $A_3B$ and $A_2B_2$ structures may
in principle be degenerate if $t$=$-8.00J_D$; the LDA value for $t$ is
smaller than this, but not unreasonably so.  In summary, the new
TB/LDA results suggest that the total orientational interactions is in
fact reasonably well described by a nearest-neighbor antiferromagnetic
Ising model, provided that current estimates of direct and indirect
energy scales are meaningful.

This work was supported in part by the Laboratory for Research
on the Structure of Matter (University of Pennsylvania), by the
NSF under the MRL program (Grant 92 20668) and by the DOE (Grant 91ER 45118).
%
%

%
%
\begin{table}
\caption{Values for the hopping constants appearing in
Eq.~(\protect\ref{tmats-yil}). To allow comparison between the two sets
of parameters, we set $t$=0.0146 eV, which equates the bandwidth of
the Gelfand/Lu spectrum with the LDA bandwidth.}
\begin{tabular}{lcccccccc}
& $A$ & $B$ & $C$ & $D$ & $E$ & $F$ & $X$ & $Y$ \\
\tableline
LDA fit    &
\makebox[.25in][r]{-0.20} &
\makebox[.25in][r]{0.10} &
\makebox[.25in][r]{-2.77} &
\makebox[.25in][r]{1.25} &
\makebox[.25in][r]{-2.74} &
\makebox[.25in][r]{1.12} &
\makebox[.25in][r]{-0.66} &
\makebox[.25in][r]{-0.32}\\
Gelfand/Lu &
\makebox[.25in][r]{0.01} &
\makebox[.25in][r]{0.38} &
\makebox[.25in][r]{-2.29} &
\makebox[.25in][r]{2.09} &
\makebox[.25in][r]{-2.36} &
\makebox[.25in][r]{0.38} &
\makebox[.25in][r]{-0.63} &
\makebox[.25in][r]{-0.49}\\
\end{tabular}
\end{table}
\begin{table}
\caption{Values for the hopping constants appearing in
Eq.~(\protect\ref{tmats-gel}).}
\begin{tabular}{lcccccccc}
& $a$ & $b$ & $c$ & $d$ & $e$ & $f$ & $g$ & $h$ \\
\tableline
LDA fit    &
\makebox[.25in][r]{-0.27} &
\makebox[.25in][r]{-2.64} &
\makebox[.25in][r]{2.36} &
\makebox[.25in][r]{-1.65} &
\makebox[.25in][r]{2.20} &
\makebox[.25in][r]{1.45} &
\makebox[.25in][r]{-3.46} &
\makebox[.25in][r]{-3.89}\\
Gelfand/Lu &
\makebox[.25in][r]{0.83} &
\makebox[.25in][r]{-1.98} &
\makebox[.25in][r]{3.36} &
\makebox[.25in][r]{-1.91} &
\makebox[.25in][r]{1.75} &
\makebox[.25in][r]{2.08} &
\makebox[.25in][r]{-3.71} &
\makebox[.25in][r]{-2.67}\\
\end{tabular}
\end{table}
\begin{table}
\caption{The three positive eigenvalues of the Hamiltonian for hopping
back and forth across a single bond, for like ($AA$) and unlike ($AB$)
orientations.}
\begin{tabular}{lccccccc}
&&  $AA$  && \makebox[.1in] &&  $AB$  & \\
\tableline
LDA fit    & 3.99 & 1.91 & 1.66 && 4.42 & 3.89 & 1.25 \\
Gelfand/Lu & 4.44 & 1.91 & 0.25 && 5.03 & 2.67 & 0.43 \\
\end{tabular}
\end{table}
\begin{table}
\caption{Kinetic energy of the conduction electron states in three different
orientational phases, in units of $t$. In the bottom panel the energies
are given relative to the ordered $A_4$ phase.  Note the overall
increase in magnitude of both relative energies in the LDA results, as
well as the reversal of the ordering between the LDA and Gelfand/Lu results.}
\begin{tabular}{lcc}
Structure & \multicolumn{2}{c}{Kinetic energy/molecule (t)} \\
          & LDA fit & Gelfand/Lu \\
\tableline
$A_4$ 		& -25.12 & -24.13 \\
$A_3B$ 		& -28.18 & -28.08 \\
$A_2B_2$	& -28.43 & -27.53 \\
\tableline
$A_3B - A_4$ 	& -4.05 & -2.94  \\
$A_2B_2 - A_4$ 	& -4.30 & -2.39
\end{tabular}
\end{table} %
%
\begin{figure}
\caption{Spectra for orientationally ordered fcc $A_3$C$_{60}$, as
calculated by self-consistent local-density methods (top), the
tight-binding parameters fit to LDA spectra (middle), and the
parameter set introduced by Gelfand and Lu (bottom).  In the middle
panel, spectra are shown for TB parameters fitted to LDA
spectra of two different orientational phases, illustrating the
transferability of these parameters. The dotted line is the Fermi energy.}
\end{figure}
\begin{figure}
\caption{Spectra for orientationally modulated $A_2B_2$ structure
described in the text. Panels are labeled as in Fig.~1.}
\end{figure}
\begin{figure}
\caption{Spectra for an ensemble average of 27-molecule supercells
with quenched disorder, using two different hopping-parameter sets.}
\end{figure}
\end{document}